\documentstyle[12pt,titlepage]{article}
\def\doublespace {\smallskipamount=7.5pt plus2pt minus2pt
                  \medskipamount=15pt plus4pt minus4pt
                  \bigskipamount=30pt plus8pt minus8pt
                  \normalbaselineskip=30pt plus0pt minus0pt
                  \normallineskip=2pt
                  \normallineskiplimit=0pt
                  \jot=7.5pt
                  {\def\smallskip {\vskip\smallskipamount}}
                  {\def\medskip   {\vskip\medskipamount}}
                  {\def\bigskip   {\vskip\bigskipamount}}
                  {\setbox\strutbox=\hbox{\vrule
                    height21.0pt depth9.0pt width 0pt}}
                  \parskip 15.0pt
                  \normalbaselines}

\def\bib{\bibitem}
\def\be{\begin{equation}}
\def\ee{\end{equation}}
\def\barr{\begin{array}}
\def\earr{\end{array}}
\def\lsim{\raisebox{-0.5ex}{$\stackrel{\textstyle<}{\sim}$}}

\def\tilm{\widetilde{m}}

\def\co{{\cal O}}
\def\ao{{\cal A}_1}
\def\at{{\cal A}_2}
\def\ce{{\cal E}}
\def\fo{{\cal F}_1}
\def\ft{{\cal F}_2}
\def\sm{{\sl SM} }
\def\mn{$\mu_\nu$ }
\def\vev{{\it v.e.v.} }
\def\vevs{{\it v.e.v.}s }
\def\im{\:{\rm Im}\:}
\def\ie{{\it i.e.} }
\begin{document}

\doublespace
\begin{center}
{\Large \bf Discrete Symmetry, Neutrino Magnetic Moment
       and the 17 keV Neutrino}   \\[7ex]
{\large \bf Debajyoti Choudhury}\footnote{Bitnet address:
         debchou@tifrvax} \\[2ex]
      {\it  Theoretical Physics Group,
        Tata Institute of Fundamental Research\\
        Homi Bhabha Road, Bombay 400 005, India.} \\[10ex]
{\bf Abstract}\\[2ex]
\end{center}
\thispagestyle{empty}

\begin{quotation}
The problem of generating  large transition magnetic moments for
nearly massless neutrinos in a truly three--generation case is
discussed. A model to achieve the same by exploiting an
octahedral symmetry is presented. The scheme also accomodates a
radiatively generated mass of $17\:keV$ for a pseudo--Dirac
neutrino that decays rapidly through the Majoron channel.
\end{quotation}

\vspace{.5in}
\vfill
\newpage

\doublespace
Two problems in neutrino physics have attracted much
attention over the past few years. The first, and relatively
longstanding one, deals with the deficiency in the solar neutrino
count in the Davis and Kamiokande experiments \cite{davis}
and the related
matter of the apparent anticorrelation between the observed
solar neutrino flux and the sunspot activity \cite{anticor}. The
other, more recent one, is
concerned with the reported signature of a $1\%$ admixture of a
$17\:keV$ neutrino with the $\nu_e$ \cite{simpson}.

While the first problem can be resolved by postulating a
relatively large magnetic moment for the neutrino \cite{vvo}, to
generate the latter in realistic models is no mean task. For
such an attempt normally leads to too large a value for the
neutrino mass. An elegant solution to this problem was suggested
by Voloshin
\cite{voloshin} in the form of a $SU(2)_\nu$ symmetry connecting
$\nu_L$ and $\nu_R$ (or $\nu_e$ and $\nu_\mu$ if you are
interested in transition moments) so that the mass term is a
triplet while the magnetic moment term a singlet. In the limit
of exact $SU(2)_\nu$ symmetry you then have the spectacle of a
non--zero \mn but a identically vanishing $m_\nu$. Several
models \cite{magmom models,our model} have been constructed
using this
idea and some variants, but most require some amount of
fine--tuning. The reason lies in the phenomenological necessity
of breaking the continuous non--abelian symmetry at a scale too
high to protect $m_\nu$ \cite{leurer}.

A way out of this imbroglio is to employ a non--abelian discrete
symmetry instead, an idea that has been richly harvested
\cite{discrete}. An aesthetic problem persists though in such
attempts, in the form of the unequal treatment they mete out to
the Standard Model (\sm) fermions. The point to remember is that
if you  put all the \sm $\nu$'s in the same representation, then
for an odd number of generations it is the mass term that
contains the singlet and not the \mn term \cite{ftn1}.
Hence, for three generations the Voloshin mechanism does not
work. Instead, one should attempt to construct models wherein
the lowest dimensional higgs operators coupling to the neutrino
current are antisymmetric in the generation index \cite{our model}.
 To achieve this in a model where the $\nu$'s lie in a
representation $R$ of the symmetry group, it is essential that
the symmetric and antisymmetric parts of $R \times R$ lie in
inequivalent irreducible representations.

In our efforts to construct a model based on such ideas, we find
that a very slight extension of the same also
affords a solution to the second problem mentioned at the outset
of this letter. Though phenomenolgical considerations
\cite{manoj} indicate that the new find is most probably a Dirac
particle and that it may be identified with the $\nu_\tau$, yet
many embarassing questions remain.
 Not the
least of which are
the questions of generating such a low scale, and, more
importantly, satisfying the strict theoretical constraints
emanating from cosmology \cite{dicus} and primordial
nucleosynthesis \cite{nucleo}. Though some models have been
proposed \cite{17models,mag17}, only one of these \cite{mag17}
makes an effort to connect the two issues that have been raised
here.

For our purpose, we choose the (24--element) symmetry group
($\co$) to be that of the
octahedron, \ie the one generated by rotations about three
4--fold axes
($f_i$), four 3--fold axes ($t_K$) and six 2--fold axes
($z_\alpha$) \cite{hamer}. The  group algebra is given by
$f_i^4 = e$, $t_1 = f_2 f_3$, $t_2 = f_3 f_1$, $t_3 = f_1 f_2$,
$t_4 = f_1 t_2 f_3$, $z_i = f_i t_i^2$, $z_{i+3} = f_i t_4^2$.
$\co$ has five
irreducible representations namely
\[
\barr{l}
   \ao :  f_i =1; \quad
   \at :  f_i = -1;  \quad
   \ce : f_1 = \sigma_1, \:
           f_2^\ast = f_3 = \left(-\sigma_1 + \sqrt{3} \sigma_2
                               \right)/2;
                \\
   \fo : f_1 = \exp(\pi T_1 /2),\: f_2 = \exp(-\pi T_2 /2),\:
           f_3 = \exp(-\pi T_3 /2); \quad
    \ft : f_i = -f_i(\fo)
\earr
\]
where $(T_i)_{jk} = \epsilon_{ijk}$. Note that only ${\cal
F}_{1,2}$ are faithful representations. The Clebsch--Gordan
decomposition is given by ($A$ and $S$ denote symmetry
properties)
\[
\barr{l}
 \fo \times \at = \ft; \quad      \ce \times \at = \ce; \quad
   \fo \times \ce = \fo + \ft;
       \\
   \fo \times \fo = \left( \ao + \ce + \ft \right)^S + \fo^A;
                      \quad
   \ce \times \ce = \left( \ao + \ce  \right)^S + \at^A;
\earr
\]
the rest following trivially.

{\bf The model: } To the standard model fermions, we
add a charge $+1$ vector singlet pair of leptons per
generation. Also we introduce three right--handed neutrino fields.
The new additions however are given an unconventional assignment
of the total lepton number, which is  conserved explicitly. The
quarks are the same as in
the \sm and we shall not talk about them any further. The entire
leptonic spectrum (under $SU(2)_L \otimes U(1)_Y \otimes \co
\otimes U(1)_l$) is then $L_L \: (2, -1/2, \fo, 1)$, $E_R\: (1,
-1, \fo, 1)$, $F_{L,R} \: (1, 1, \fo, 1)$,
 $N_{1R} \: (1, 0, \ao, -1)$, $N_{2R} \: (1, 0, \ao, -2)$  and
  $N_{3R} \: (1, 0, \ao, -4)$

As for the scalar sector, apart from the $\phi \: (2, 1/2, \ao,
0)$ and $H \: (2, 1/2, \ce, 0)$ which give masses to the \sm
fermions, we also have
$\Sigma \: (1, 0, \fo, 5)$ and $\sigma \: (1, 0, \ao, 6)$ to break the
lepton number and give a Majorana mass term;
$\Omega \: (1, 1, \fo, 7)$, $\Xi \: (1, 1, \ao, 6)$,
$\chi \: (2, 3/2,\ao, 0)$ and
$\eta \:(2, 1/2, \fo, 2)$ that traverse in loops
responsible for various radiative generations; and finally
$\xi \:(2, 1/2, \fo, -2)$ and $\zeta \:(2, 1/2, \fo, -3)$ to
give Dirac masses to the
neutrinos.

The fermion mass and Yukawa terms then read
\be
\barr{rcl}
{\cal L}_{m + Y} & =& \tilm \overline{F_L} F_R +
                      \overline{L_L} E_R (a_1 \phi + a_2 H)
            + b_1 \overline{N_{1R} }L_L \xi
            + b_2 \overline{N_{2R} }L_L \zeta \\
        & & + c \overline{N_{2R}^c} N_{3R} \sigma
            + g_1 \overline{F_R} L_L \chi
            + g_2 \overline{L_L^c} F_L \eta^\dagger + H.c.,
\earr
\ee
while the higgs potential, apart from the usual quadratic and
quartic invariants, also contains the cross terms
\be
\barr{rcl}
{\cal L}_{\rm Higgs} &=&
         \Omega^\dagger \Sigma \eta
                  \left(\lambda_1 \phi + \lambda_1' H \right)
        + \lambda_2 \Xi^\dagger \sigma^\dagger \Omega \Sigma
        + \lambda_3  \chi^\dagger \sigma^\dagger \Xi \phi
                  \\
     & & + \lambda_4 \chi^\dagger \Sigma^\dagger \xi \Omega
       + \lambda_5 \zeta^\dagger \sigma^\dagger \xi \Sigma
         + \mu_1 \eta^\dagger \zeta \Sigma
         + \cdots           \\
\earr
\ee
where we have displayed only those terms that interest us. In
all of above the Clebsch--Gordan coefficients are implicitly
present.

The fields $\eta,\zeta,\xi$ are ascribed a positive ${\rm (mass)}^2$
value so that they do not gain a vacuum expectation value (\vev)
at the tree level. One good feature of our model is that we do
not need to introduce a new high scale as all symmetry including
$\co$ and the lepton number are broken at the weak scale. The
tree--level \vevs are then
\be
    \langle \sigma \rangle = s ; \quad
    \langle \Sigma \rangle = (S_1, S_2, S_3) ; \quad
    \langle \phi \rangle = v_s ; \quad
    \langle H \rangle = (v_1, v_2) ;
\ee
where only the $\co$ dependence is exhibited. Apropos the domain
wall problem, it can be tackled \cite{domain walls} by either
invoking symmetry non--restoration in multi--Higgs models or the
possible absence of high temperature phase transition in a
system with large net lepton number as is the case here.

To this level then, the charged lepton mass matrix
is diagonal with all three exotic particles
degenarate with a mass $\tilm \sim 200 \: GeV$. This form
assures that there are no flavour changing neutral currents
(FCNC) to the leading order. The model however cannot explain
the \sm fermion mass hierarchy which is to be taken care of by
appropriate choice of \vevs and Yukawa couplings. On the
other hand, no Dirac masses for the neutrinos have been
generated and the
neutrino mass matrix is of rank two.

A magnetic moment for
the neutrino is generated through the diagram in Figure 1 on
insertion of a photon on either internal line. The contribution
to \mn can be symbolically expressed as
\be
   \mu_\nu \sim \displaystyle
      \frac{2 e}{16 \pi^2}
      \frac{g_1 g_2 \lambda_1 \lambda_2 \lambda_3 S^2 s^2 v^2}
           {\tilm^7 (x_\chi - x_\Omega)}
        \left[ \frac{ h(x_\Xi, x_\eta) - h(x_\Xi, x_\chi)}
                    { x_\eta - x_\chi }
             - \frac{ h(x_\Xi, x_\eta) - h(x_\Xi, x_\Omega)}
                    { x_\eta - x_\Omega }
         \right]
                  \label{mu-nu}
\ee
where
\[
         h(x,y) = \displaystyle \frac{f(x) - f(y) }
                                     { x - y },
\]
$f(x) = (1 - x)^{-3} \left[ (1 -4x + 3 x^2)/2 - x^2 \ln x
\right]$ and $x_\chi \equiv m_\chi^2/\tilm^2 $. The function
$f(x)$ is monotonically decreasing with $f(0) = 1/2$, $f(1) =
1/3$ and $f(\infty) = 0$. It should be noted that the above
is only the contribution for a particular set of fields
travelling in
the loop. The full family dependence of \mn can easily be
obtained by summing over all such diagrams taking into account
the different masses, \vevs and couplings. To get an order of
magnitude estimate, we assume that all the scalars and the
$F$--fields have mass $\sim O(200\:GeV)$ and that the couplings
in eqn(\ref{mu-nu}) are each $O(0.1)$. We then have
\be
    \mu_\nu \sim O \left( 10^{-11} \mu_B \right)
       \label{mu-val}
\ee
and hence of the correct order of magnitude to explain the
observed anticorrelation \cite{anticor,vvo}.

Normally, with the  removal of the
photon, this diagram would generate a mass correction for the
neutrino thus requiring fine--tuning. However, in the present
model, this correction term is antisymmetric in the generation
index and hence does not contribute at all to the neutrino
Majorana mass. As pointed out right at the beginning, this is {\sl not}
a consequence of a Voloshin--like symmetry. Rather, unlike in the
Voloshin mechanism, here the \mn term is not a group invariant
and hence cannot arise until after the symmetry is broken. The
key to the protection of the mass lies in structure of the
theory and more particularly that of the lowest--order diagram
leading to \mn. A look at the fermion line of Figure 1 shows
that irrespective of the scalars traversing the loop, the
effective operator coupling to the neutrino current has to be
antisymmetric. This result owes its origin to the fact that the
mass term for the exotic fermions ($F_{L,R}$) is
$\co$--invariant (\ie independent of the breaking) and hence
proportional to the unit matrix in
the generation space. Any departure from such structure is
caused only by higher--dimensional operators and shall be
commented upon later.

This would
have been the whole story were it not for the fields $N_{iR}$
and the scalars $\xi$ and $\zeta$. Though there are no three or
four--dimensional operators leading to \vevs for them,
higher--dimensional operators arising from radiative corrections
do contribute to $\langle \xi_i \rangle $ {\it etc}.  A typical
example is the operator $\xi \Sigma^{\dagger 2} \sigma^2 \phi$
(as in Figure 2), resulting in
\be
   \langle \xi_i \rangle \sim \displaystyle
                \frac{\lambda_2  \lambda_3 \lambda_4}{16 \pi^2}
                \frac{ S^2 s^2 v}{m^2_\xi m^2_{\rm loop}}  \sim
                O (1 \:MeV).
\ee
where $m_{\rm loop}$ is the typical mass of the scalars in the
loop.  Similar values for $\langle \zeta_i \rangle$ and $\langle
\eta_i \rangle$ are also generated through such diagrams and
mixings with each other. Non--zero $\langle \eta_i \rangle$  of
course lead to mixings of the \sm charged leptons with the
exotics, but due to the huge disparity in scales the levels of
FCNC are somewhat below current experimental limits. The
neutrino mass matrix, in the ($\nu_i \:N_1 \:N_2 \:N_3$)
basis (where $\nu_i$ represent the \sm particles and all fields are of the
same helicity), now reads
\be
         M_{\nu}  =\pmatrix{0 & M_1^T \cr
                            M_1 & M_2 \cr}, \quad
        M_1 \equiv \pmatrix{\alpha_1 & \alpha_2 & \alpha_3 \cr
                            \beta_1  & \beta_2  & \beta_3 \cr
                                  0  &       0  &       0 \cr},  \quad
        M_2 \equiv \pmatrix{0 & 0 & 0 \cr
                            0 & 0 & M \cr
                            0 & M & 0 \cr}
             \label{one loop mass}
\ee
where $\alpha_i = b_1 \langle \xi_i \rangle $,
 $\beta_i = b_2 \langle \zeta_i \rangle $ and $M = c s$.
$M_\nu$, which is of rank 4, has the eigenvalues $0$, $0$,
$\pm \left[ \left( G - \sqrt{G^2 - 4 H} \right) /2 \right]^{1/2}$ and
$\pm \left[ \left( G + \sqrt{G^2 - 4 H} \right) /2 \right]^{1/2}$.
Here $G = M^2 + {\vec \alpha}^2 + {\vec \beta}^2$ and
$H = M^2  {\vec \alpha}^2 +
        \left({ \vec \alpha} \times {\vec \beta}\right)^2$.
Note that $\alpha_i, \beta_i$ can naturally be $\sim O(10\:keV)$
without requiring either an artificial generation of such a
scale or unnaturally small Yukawa couplings. Assuming $M \sim
250\:GeV$, the neutrino spectrum then consists of three apparently--Dirac
particles --- one superheavy, one massless and one of mass
$17\:keV$.  The mixing of $\nu_{17}$ with $\nu_e$ is engendered
by the ratios of the Dirac mass terms and easily give the
required strength.

At this stage it is as well to point out that the full symmetry
of $M_\nu$ is not a symmetry of the theory and hence is broken
by quantum corrections. For example, the off--diagonal mass
terms for the charged leptons arising out of $\langle \eta_i
\rangle$ would lead to non---trivial mixing in that sector and
hence to neutrino mass corrections through diagrams as in Fig.
1. However, due to the smallness of $\langle \eta_i \rangle$,
these corrections are almost of the see--saw type in magnitude
($ \sim 10^{-3} \:eV$) and do not alter the neutrino spectrum to
any significant degree. Also, higher loop diagrams generate
Majorana mass terms of similar order and involving ``ordinary''
neutrinos. As a result of all these, the mass degeneracies are
lifted and the Dirac neutrinos split into three pairs of
pseudo--Dirac particles. The small masses for $\nu_e$ and
$\nu_\mu$ that are thus generated would be adequate for a MSW
type of resonance enhancement in the Sun \cite{akhmedov}.
Also the
effective mass contributing to the neutrinoless double beta decay [ $\beta
\beta_{0 \nu}$], is $\lsim \beta_1^2/M$ and though miniscule, affords an
example
where the effective Majorana mass for $\beta \beta_{0 \nu}$
could be  larger than
that to be observed in Kurie plots \cite{double beta}.

Of course, one might wonder if diagrams analogous to those in
Fig. 1, but with $N_{iR}$ as the virtual leptons instead of
$F_{L,R}$ would contribute to Majorana mass terms. For if they
did, the earlier group theoretic argument leading to exact
cancellations would not hold and indeed the contributions could be
large. However, it is easy to see that there is no place for
such apprehension. Two facts need to be noted. Firstly, there is
no Dirac term involving $N_{3R}$ and secondly, the only tree
order (and hence large) Majorana mass term is of the form
($  \overline{N_{2R}^c} N_{3R}  + H.c. $). As a result, there
can exist no one--loop diagram with $N_{iR}$ as the internal
particle(s) and contributing to the neutrino Majorana masses. This
can be verified rigorously by working with the mass eigenstates
instead. Such arguments obviously do not hold for complicated
multi--loop diagrams, but those contributions are too small to
be relevant.

The Majoron (the only surviving Goldstone boson in the theory),
to the leading order, is given by
\be
\vartheta \sim \left(
        6 s \im \sigma + 5 S_i \im \Sigma_i
        + 2 \langle \eta_i \rangle \im \eta_i
        - 2 \langle \zeta_i \rangle \im \zeta_i
        - 3 \langle \xi_i \rangle \im \xi_i
                   \right)/N
\ee
(where $N$ gives the normalization) and is hence primarily a
$SU(2)_L$ singlet. Thus its coupling with the \sm charged
leptons is highly suppressed and fully in consonance with the
bounds coming from $Z$--decay width \cite{zdecay} as well as
astrophysical considerations\cite{kim}. However, if one
considers the coupling of the $\nu$'s with the Majoron, one gets
\be
   G_{\nu \vartheta} \approx
      N^{-1}   \pmatrix{0 & G_1^T \cr
                  G_1 & 6M_2 \cr},      \quad
        G_1 \equiv \pmatrix{2\alpha_1 & 2\alpha_2 & 2\alpha_3 \cr
                            3\beta_1  & 3\beta_2  & 3\beta_3 \cr
                                   0  &         0 &        0 \cr},
             \label{majoron coupling}
\ee
which is not diagonalized simultaneously alongwith $M_\nu$.
This then leads to a nondiagonal $\nu - \vartheta$ coupling of the
order of $m_\nu /N$
and as a
consequence to a very fast decay of the $17\:keV$ neutrino which
would have a lifetime $\sim O (10^5\:sec)$.

   To conclude, we have presented a model based on a
non--abelian discrete symmetry $\co$ that leads to a significant
amount of transition magnetic moment for nearly massless
neutrinos. The model is {\it not} a discrete version of the
Voloshin mechanism, which we have argued cannot work for the
truly 3--generation case. Rather, the protection of $m_\nu$ owes
its existence to the absence of any family--symmetric effective
scalar operator to the lowest order. The magnetic moment term
itself arises on breaking the symmetry, which, being discrete,
can be preserved until at least the weak scale. Higher order
effects do lead to small mass corrections but these are greatly
suppressed.

A simple extension of this model is shown to accomodate a
pseudo--Dirac
$17\:keV$ neutrino as well. The latter can be identified with
the $\nu_\tau$ and is generated through a
cripple see--saw mechanism that keeps $\nu_e$ and $\nu_\mu$
massless.
 However, tiny FCNCs in the charged lepton
sector and multiloop diagrams together cause small mass corrections of the
order of $10^{-3}\:eV$.  The $\nu_{17}$ decays very fast into a lighter
neutrino
and a singlet--doublet Majoron and is thus consistent with all
known experiments, whether terrestrial or cosmic.

\noindent
{\bf Acknowledgements}
The author would like to thank Utpal Sarkar for useful
discussions and suggestions. Thanks are also due to the referees for
illuminating comments.

\newpage
\noindent
{\bf Figure Captions}

{\bf Figure 1.} Diagrams (sans photon lines) contributing to
neutrino magnetic moments.

{\bf Figure 2.} Typical diagram leading to radiative generation of
$\langle \xi_i \rangle$.
\end{document}